\documentclass[12pt,eqsecnum,preprint]{aastex}
\title{X-ray Absorption in Type II Quasars: Implications for the Equatorial Paradigm of
Broad Absorption Line Quasars}
\author{Brian Punsly}
\affil{4014 Emerald Street No.116, Torrance CA, USA 90503 and
International Center for Relativistic Astrophysics,
I.C.R.A.,University of Rome La Sapienza, I-00185 Roma, Italy}
\email{brian.m.punsly@L-3com.com or brian.punsly@gte.net}
\begin{document}
\begin{abstract}In this article, the hydrogen column densities derived from
X-ray observations of type II (hidden) quasars and broad
absorption line quasars (BALQSOs) are compared. These column
densities represent the amount of absorbing material between the
X-ray source and the observer. A sample of type II QSOs with
strong narrow emission lines, weak UV continuum and no broad
emission lines that also have deep X-ray observations was
collected from the literature for analysis. The standard model of
equatorial BAL (broad absorption line) winds predicts that the
column densities of this type II QSO sample should significantly
exceed BALQSO column densities. Based on the existing published
deep hard X-ray observations this does not seem to be true, the
BALQSO absorption columns are anomalously large. Actually, the
limited existing data indicate that BALQSOs have column densities
which are larger than the type II QSO column densities at a
statistically significant level. The implication to BAL outflows
and the fundamental physical geometry of QSOs is discussed.
\end{abstract}
\section{Introduction} A popular paradigm for broad absorption line quasars
(BALQSO) is that the UV broad absorption lines (BALs) form in an
equatorial wind that flows just above the edge of a dusty
molecular torus. This notion is a refinement of the what is called
the "standard model for AGN," consisting of a hot accretion flow
onto a black hole and a surrounding torus of molecular gas
\citep{ant93}. It is widely believed that all radio quiet quasars
have BAL outflows, but the designation of a quasar as a BALQSO
depends on whether the line of sight intersects the solid angle
subtended by the outflow. Within this extension of the standard
model of AGN, along lines of sight to the accretion disk that
approach from high latitudes, within $\sim 40 ^{\circ}$ of the
pole, one views a conventional quasar spectrum without BALs, a
type I AGN. About 20\% of QSOs show BAL outflows, loosely defined
as UV absorbing gas that is blue shifted at least 5,000 km/s
relative to the QSO rest frame and displaying a spread in velocity
of at least 2,000 km/s \citep{wey97,hew03}. For lines of sight to
the accretion disk at intermediate latitudes, say $\sim 40^{\circ}
- 50^{\circ}$, one classifies the object as a BALQSO since a
quasar spectrum is seen except for the the blue sides of the UV
broad emission lines (BEL) which are absorbed, i.e., the near side
of the BEL is obscured. At lines of sight even closer to the
equatorial plane, $> 50^{\circ}$, one encounters the dusty torus,
the quasar is completely obscured; both the red and blue side of
the BELs are attenuated and the central continuum optical/UV
source is hidden from view, these are known as type II quasars.
Consequently, in general, there is a larger attenuating column
along lines of sight to the active nucleus for type II quasars
than BALQSOs, if the paradigm is correct.
\par  We check this prediction by comparing the column densities toward the nuclear X-ray source
measured by XMM and Chandra in an archival sample of spectral
 fits in the existing literature on BALQOs and type II QSOs. This has only
 recently been made possible. In the past 5 years, some high quality bona fide BALQSOs
 data has been reduced and published ($\sim 13$) and a somewhat larger sample of quality
 type II QSO X-ray spectra ($\sim 33 $)
has appeared in the literature. In this article, we define type II
QSOs (QSOIIs, hereafter) as strong narrow emission line objects
with a weak UV continuum, no broad emission lines and a hidden
nucleus of quasar luminosity that is verified by IR or X-ray
observation. Contrary to the expectations of the the equatorial
wind model of BALQSOs, preliminary data indicate that QSOIIs
actually have a distribution of X-ray absorbtion hydrogen column
densities that are statistically smaller (at the $ 99.8\%$
confidence level by a Mann-Whitney test) than the distribution of
X-ray absorption columns measured for BALQOs. It is likely that
this result is largely sample selection dependent, but the weaker
claim that there is no statistical support for the hypothesis,
QSOII X-ray absorption column densities are larger than BALQSO
X-ray absorption column densities, seems well justified by the
observations to date.
\par The data is intriguing even though it is based on small samples and
certainly deserves further study. The implication, if the trend
holds with more data, is that there is an additional absorbing
column that obscures the X-ray emission in some or all BALQSOs
that is not encountered for low latitude lines of sights
characteristic of QSOIIs. In the equatorial wind model, this extra
absorber would be the so-called "hitchhiking gas" that shields the
BAL wind from the X-ray source \citep{mur95}. However, this is by
necessity located at low latitudes and should be encountered by
most if not all type II lines of sight. By contrast polar BAL wind
models, naturally produced an additional X-ray absorber along the
polar axis, so this extra absorber would not be detected in QSOIIs
\citep{pun99,pun00}. The comparison of column densities in QSOIIs
and BALQSOs is a vital clue for understanding the fundamental
underlying physical configuration that constitutes a QSO.
\section{The Deep X-ray Sample} In order to accurately
determine large X-ray absorbing columns,
$N_{H}>10^{22}\mathrm{cm}^{-2}$, high energy (10-20 keV in the
quasar rest frame) observations are required. This essentially is
a consequence of the fact that at $N_{H}=10^{23}\mathrm{cm}^{-2}$
the absorbing column is virtually black to soft X-rays
\citep{mur95}. One needs higher energy X-ray measurements to
determine the continuum spectral index, so the intrinsic
background flux can be extrapolated back to the soft X-ray regime.
In \citet{pun05} anecdotal examples of misleading observations of
BALQSOs that were used to estimate $N_{H}$ based on very few or no
hard photons were given. The most notable examples are MRK 231 for
which the column density was under estimated by two orders of
magnitude for years, \citet{bra04}, and PHL 5200 which was under
estimated by over 3 orders of magnitude, \citet{bri02}. Thus, the
ROSAT soft X-ray observations of \citet{gre96} are inadequate for
restricting the column density. Also, the survey work of
\citet{gal06} with $\sim 4\pm 3$ detected hard photons for the
vast majority of the sources is unsuitable for this analysis. The
lessons learned form past experience is that there are no shortcuts
 to attaining even moderately accurate estimates of the X-ray
absorption columns in BALQSOs. These are necessarily difficult
observations that require very long exposures. Secondly, the
fluxes from an obscured nucleus can be so attenuated that it can
be swamped by any nearby weak X-ray source \citep{bri02}. Thus,
high resolution is preferred. These limitations indicate that
reliable results can only be obtained from archival Chandra and
XMM data for samples of highly attenuated sources such as QSOIIs
and BALQSOs.
\par An important clarification of the
BALQSO sample is that these are bona fide BALQSOs, in particular
the so-called "mini-BALQSOs" are excluded (such as RXJ
0911.4+0551, \citet{mor01} and PG 1115+080 \citet{cha03}). This
class of sources has been preferentially proposed for observation
as a premeditated expedience to get X-ray bright objects that are
BALQSO-like without the risk of low photon counts inherent to
BALQSOs observations \citep{cha03}. Thus, they are likely to
contaminate the statistics of the small BALQSO sample with their
lower absorption columns.
\begin{table}
\caption{X-ray Absorption Column Densities of BAL and type II
QSOs}
\begin{tabular}{ccccc}
\tableline \rule{0mm}{3mm}
\footnotesize Source & \footnotesize z & \footnotesize type & \footnotesize Column Density& \footnotesize X-ray reference\\
       & &   & \footnotesize ($10^{23}\mathrm{cm}^{-2}$)  &  \\
\tableline \rule{0mm}{3mm}
\footnotesize LBQS 2212-1759& \footnotesize  2.217    & \footnotesize BAL &  \footnotesize 250 &  \footnotesize \citet{cla06}\\
\footnotesize MRK 231 & \footnotesize   0.042     &  \footnotesize BAL &  \footnotesize $26.5^{+17.3}_{-8.5} $ & \footnotesize \citet{bra04}\\
\footnotesize PHL 5200 & \footnotesize   1.98 &  \footnotesize BAL & \footnotesize $>10$  &  \footnotesize \citet{bri02,mat01}\\
\footnotesize 1413+117 & \footnotesize 2.56 & \footnotesize BAL & \footnotesize $> 10^{a}$  & \footnotesize \citet{cha04,osh01}\\
\footnotesize CDF-S 202 & \footnotesize 3.700 & \footnotesize II &  \footnotesize $> 10^{a}$   & \footnotesize \citet{nor02,toz06}\\
\footnotesize CDF-S 153 & \footnotesize  1.54 & \footnotesize II &  \footnotesize $> 10^{a}$   & \footnotesize \citet{toz06}\\
\footnotesize IRAS 09104+4109 & \footnotesize 0.442 & \footnotesize II &  \footnotesize $> 10^{a}$   &  \footnotesize \citet{iwa01}\\
\footnotesize IRAS 07598+6508 & \footnotesize  0.148 & \footnotesize BAL &  \footnotesize $> 10^{b}$   &  \footnotesize \citet{ima04}\\
\footnotesize PG 0946+301 & \footnotesize  1.233 &  \footnotesize BAL &  \footnotesize 10  & \footnotesize \citet{mat00}\\
\footnotesize SDSSJ 1226+0131 & \footnotesize 0.732 & \footnotesize II &  \footnotesize $8.7^{+6.3}_{-2.6}$  &  \footnotesize \citet{pta06}\\
\footnotesize CDF-N 171 & \footnotesize 1.993 & \footnotesize II &  \footnotesize $ 6 $   &  \footnotesize \citet{ale05}\\
\footnotesize CDF-N 190 & \footnotesize 2.005 & \footnotesize II &  \footnotesize $ 5 $   &  \footnotesize \citet{ale05}\\
\footnotesize CXO 52 Lynx & \footnotesize 3.288  & \footnotesize II & \footnotesize 5 &  \footnotesize \cite{stu06}\\
\footnotesize PG 1254 + 047 & \footnotesize 1.024 & \footnotesize BAL & \footnotesize $2.8\pm 0.3^{c}$  & \footnotesize \citet{sab01}\\
\footnotesize CDF-S 27 & \footnotesize 3.064 & \footnotesize II &  \footnotesize $ 2.8^{+0.92}_{-0.8} $   &  \footnotesize \citet{toz06}\\
\footnotesize CDF-S 51 & \footnotesize 1.097 & \footnotesize II & \footnotesize $2.2^{+0.29}_{-0.24}$   & \footnotesize \citet{toz06}\\
\footnotesize SDSSJ 0801+4112 0.556 & \footnotesize 0.556 & \footnotesize II & \footnotesize $1.6^{+1.1}_{-0.63}$ & \footnotesize \cite{pta06}\\
\footnotesize CDF-S 54 & \footnotesize 2.561 & \footnotesize II &  \footnotesize $ 1.07^{+0.54}_{-0.46} $   &  \footnotesize \citet{toz06}\\
\footnotesize H 05370043 & \footnotesize 1.797 & \footnotesize II &  \footnotesize $ 1.05^{+0.10}_{-0.05} $   &  \footnotesize \citet{per04}\\
\footnotesize SBS 1524 + 517 & \footnotesize 2.85 & \footnotesize BAL & \footnotesize $1.0\pm 0.3^{d}$ & \footnotesize \citet{she06}\\
\footnotesize APM 08279+52550.6 & \footnotesize 3.91 & \footnotesize BAL & \footnotesize $0.98\pm 0.23$ & \footnotesize \citet{cha02,has02}\\
\footnotesize H 15800019  & \footnotesize 1.957&  \footnotesize II &  \footnotesize $ 0.76^{+1.17}_{-0.55} $   &  \footnotesize \citet{per04}\\
\tableline{\rule{0mm}{3mm}}
\end{tabular}\\
\tablenotetext{a}{best fit has hard X-rays seen in reflection}
\tablenotetext{b}{very weak X-ray source, the primary X-ray source
is behind a Compton thick screen}
 \tablenotetext{c}{only 47 detected photons, a strong candidate for a much large column density with deeper observations} \tablenotetext{d}{based on the best fit with an ionized
absorber}
\end{table}
\begin{table}
\caption{X-ray Absorption Column Densities of BAL and type II QSOs
(continued)}
\begin{tabular}{ccccc}
\tableline \rule{0mm}{3mm}
\footnotesize Source & \footnotesize z & \footnotesize type & \footnotesize Column Density & \footnotesize X-ray reference\\
       & &   & \footnotesize ($10^{23}\mathrm{cm}^{-2}$)  &  \\
\tableline \rule{0mm}{3mm}
\footnotesize PG 2112+059 & \footnotesize 0.466 & \footnotesize BAL & \footnotesize $0.7^{+0.04}_{-0.03}$& \footnotesize\citet{gal04}\\
\footnotesize 1246-057  & \footnotesize  2.24 & \footnotesize BAL &  \footnotesize $\approx 0.7$ & \footnotesize \citet{gru03}\\
\footnotesize H 05370123 & \footnotesize  1.153& \footnotesize II &  \footnotesize $ 0.66^{+2.16}_{-0.41} $   &  \footnotesize \citet{per04}\\
\footnotesize SBS1542+241 & \footnotesize 2.361 & \footnotesize BAL & \footnotesize 0.5 & \footnotesize \citet{gru03}\\
\footnotesize LH 12A & \footnotesize  0.99 & \footnotesize II & \footnotesize $0.51\pm 0.14$ & \footnotesize \cite{stu06}\\
\footnotesize XBS J0216-0435 & \footnotesize 1.987 & \footnotesize II &  \footnotesize $0.47^{+0.15}_{-0.13}$   &  \footnotesize \citet{sev06}\\
\footnotesize SDSS J1641+3858 & \footnotesize  0.596 & \footnotesize II &  \footnotesize $0.46^{+0.41}_{-0.13}$ & \footnotesize \citet{pta06}\\
\footnotesize LH 14Z & \footnotesize 1.38 & \footnotesize II & \footnotesize 0.4 & \footnotesize \cite{stu06}\\
\footnotesize UM425 & \footnotesize 1.465 & \footnotesize BAL & \footnotesize $0.38\pm 0.12^{e}$ & \footnotesize \citet{ald03}\\
\footnotesize CDF-S 117 & \footnotesize  2.573 & \footnotesize II \footnotesize & $0.31^{+0.18}_{-0.17}$&  \footnotesize \citet{toz06}\\
\footnotesize XBSJ204043.4-004548 & \footnotesize 0.615 & \footnotesize II &  \footnotesize $ 0.30^{+0.12}_{-0.09} $   &  \footnotesize \citet{cac04}\\
\footnotesize XBSJ013240.1-133307 & \footnotesize 0.562 & \footnotesize II &  \footnotesize $ 0.25^{+0.07}_{-0.06} $   &  \footnotesize \citet{cac04}\\
\footnotesize H 50900013  & \footnotesize 1.261 & \footnotesize II &  \footnotesize $ 0.252^{+0.46}_{-0.22} $   &  \footnotesize \citet{per04}\\
\footnotesize LH H57 & \footnotesize  0.792 & \footnotesize II &  \footnotesize 0.25 & \footnotesize \cite{stu06}\\
\footnotesize SDSS J0210-1001 & \footnotesize 0.540 & \footnotesize II & \footnotesize $0.23^{+0.32}_{-0.19}$ & \footnotesize \citet{pta06}\\
\footnotesize AXJ0144.9-0345 & \footnotesize 0.620 & \footnotesize II &  \footnotesize $ 0.21^{+0.02}_{-0.04} $   &  \footnotesize \citet{geo06}\\
\footnotesize CDF-S 18 & \footnotesize 0.979 & \footnotesize II &  \footnotesize $0.19^{+0.02}_{-0.02}$   &  \footnotesize \citet{toz06}\\
\footnotesize CDF-S 31 & \footnotesize 1.603 & \footnotesize II &  \footnotesize $0.18^{+0.04}_{-0.02}$   &  \footnotesize \citet{toz06}\\
\footnotesize H 0537011a & \footnotesize 0.981 & \footnotesize II &  \footnotesize $ 0.13^{+0.15}_{-0.09} $   &  \footnotesize \citet{per04}\\
\footnotesize H 05370016 &\footnotesize 0.995 & \footnotesize II &  \footnotesize $ 0.13^{+0.16}_{-0.09} $   &  \footnotesize \citet{per04}\\
\footnotesize LH 28B & \footnotesize 0.205 & \footnotesize II &  \footnotesize 0.02 & \footnotesize \cite{stu06}\\
\footnotesize CDF-N 77 & \footnotesize 2.416 & \footnotesize II &  \footnotesize $ 0.001 $   & \footnotesize \citet{ale05}\\
\footnotesize H 03120006 & \footnotesize 0.680 & \footnotesize II &  \footnotesize $ <0.16 $   &  \footnotesize \citet{per04}\\
\footnotesize H 50900001 & \footnotesize 1.049 & \footnotesize II &  \footnotesize $ <0.11$   &  \footnotesize \citet{per04}\\
\tableline{\rule{0mm}{3mm}}
\end{tabular}\\
\tablenotetext{e}{an equally acceptable fit of $1.0\pm 0.15$ is
obtained with a warm absorber}
\end{table}
\par Over the past five years a sample of 13 BALQSOs with
suitable X-ray observations has appeared in the literature (see
table 1). Only recently, observations of QSOIIs, of similar
quality,  have been published
\citep{ale05,bel06,eva06,pta06,stu06,toz06,per04}. Thus, a
comparison of BALQSO and QSOII absorbing columns is possible for
the first time.
\par The QSOII observations are of two types, radio
loud and radio quiet. The radio loud (radio loud type II quasars
are usually referred to as narrow line FR II radio galaxies) X-ray
observations of \citet{bel06,eva06} have a resolution of only
$\sim 1$ kpc for Chandra observations and an order of magnitude
worse for XMM. Thus, contamination of the accretion flow X-ray
emission from synchrotron and inverse Compton emission from the
base of the radio jet makes the interpretation of this data, for
the purposes of this study, extremely suspect. It is curious that
the inclusion of these sources does not statistically alter the
conclusions of this article. The sample of QSOIIs in this article
is chosen to be radio quiet with spectroscopically determined
redshifts. The quasar nature of the hidden nucleus is established
by IR, X-ray and narrow line luminosity.
\par Table 1 is compiled from QSOIIs
observed with Chandra and XMM that appear in the literature with
sufficient sensitivity to detect the hard X-ray continuum (i.e.,
typically with more than 100 total counts from 0.5 keV to 8 keV).
We make the distinction of \citet{pta06} to exclude sources with a
de-absorbed (intrinsic) X-ray luminosity below the Seyfert 1/ QSO
dividing line, i.e $L_{X}> 10^{44} \mathrm{ergs/sec}$, from 0.5
keV to 10 keV. It is crucial to segregate QSOIIs from Seyfert 2
galaxies, because Seyfert galaxies are not the parent population
for BALQSOs. Although Seyfert 1 galaxies have absorption outflows,
the terminal velocities are much lower than BALQSOs \citep{wey97}.
The phenomenon of BALs is associated with the high luminosity AGN,
QSOs. Thus, in this context it is very important to make the
distinction between QSOs and Seyfert galaxies.
\par The results of
the best fit column densities are tabulated in table 1 and figure
1. All of the data in table 1 is taken from Chandra or XMM
observations except for PG 0946+301. This was a deep pointed
observation with ASCA from \citet{mat00} that was extremely long,
100ks. It is possible that a deeper observation with XMM or
Chandra would produce a better spectrum or resolve a nearby weak
background X-ray source that would indicate a larger value of
$N_{H}$ than the one in the table.
\begin{figure}
\plotone{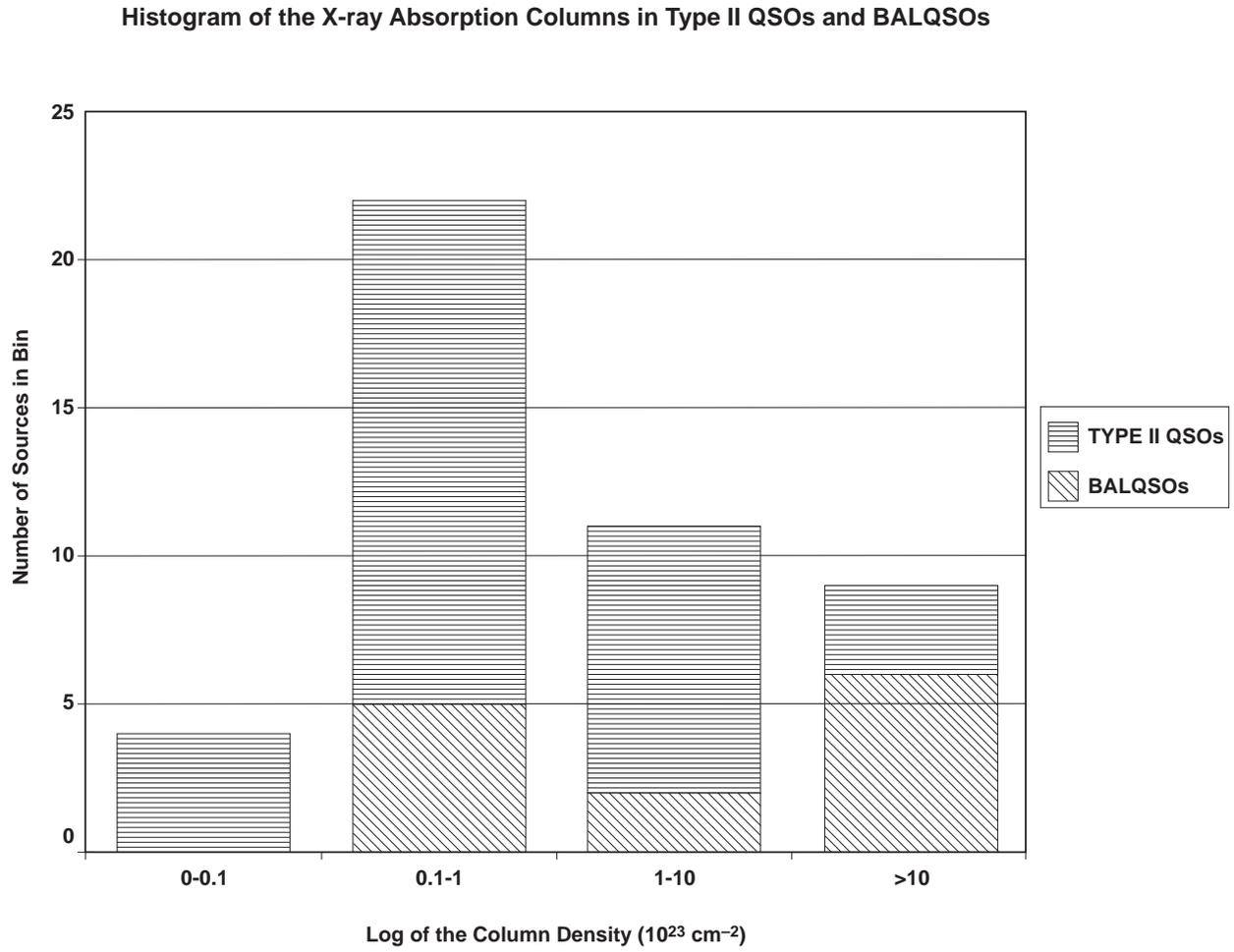}
    \caption{A histogram of the distributions of hydrogen column densities from the best fits of
    the data from deep X-ray observations}
\end{figure}
\section{Selection Effects}The sample in table 1 is biased by many selection
effects which we discuss below. \subsection{Selection Effects in
the QSO II Sample}The QSOIIs in \citet{stu06} were selected on the
basis of X-ray detections by XMM with high intrinsic absorption
(an average $N_{H}\sim 10^{23}\mathrm{cm}^{-2}$). Thus, this
method selects against low intrinsic absorption sources, $N_{H} <
10^{22}\mathrm{cm}^{-2}$ and Compton thick sources, $N_{H} >
10^{24}\mathrm{cm}^{-2}$, which might not be X-ray detections
(except in the cases in which partial covering or scattered X-rays
are significant). The sample picks X-ray detections from the 1
Msec observation of the Lockman hole with XMM, with red colors and
ISCOCAM detections. All the sources also have strong narrow lines,
confirming the hidden quasar inside and making the existence of a
completely hidden BEL very likely. Similarly, the 1Ms Chandra Deep
Field South QSOII sample of \citet{toz06} requires $N_{H} >
10^{22}\mathrm{cm}^{-2}$ and has the same selection bias as
\citet{stu06}. The cutoff for quality spectral fits in this sample
was determined by \citet{toz06} to be 120 counts, we lower this
limit down to 100 counts in order to help offset the selection
bias against highly absorbed sources in this flux limited sample.
Many of the sources in table 1 come from the HELLAS2XMM survey,
which are designated by the prefix H in table 1 \citep{per04}.
Again, this sample selects against Compton thick sources. A few
Chandra Deep Field North sources have been found in \citet{ale05}
with $>100$ counts and a spectral identification in \citet{cha05}
as an AGN based on narrow line emission and without broad line
emission. Note our selection criteria excludes the sources in
\citet{ale05} that have hidden quasars based on X-ray
observations, but do not have the emission line signature
indicative of an AGN in \citet{cha05}. Supplementing this is the
sample of \citet{pta06} which is selected from the SDSS survey of
sources without a strong UV continuum, no broad emission lines,
but strong narrow lines. This method does not select against
Compton thick absorption a priori, but in practice observations of
highly absorbed sources will lack enough counts to adequately fit
a spectrum. The X-ray observations of QSOIIs in \citet{pol06} have
too few photons to produce a reliable spectral fit for our
purposes.
\par Using the selection criteria described in the last section, some sources from \citet{stu06}
have been excluded from table 1, LH 901A is a Seyfert II galaxy
and AX J0843+2942 is radio loud. Sources were also excluded from
the sample of \citet{pta06}. Based on O III luminosity,
\citet{pta06} classify SDSS J0115+0015 and SDSS J 0243 + 0006 as
Seyfert II galaxies. The sources SDSS J0801+4412, SDSS J0842+3625
and SDSS J1232+0206 all have exposures of less than 10 ks and the
observations were unable to collect enough photons to constrain
the X-ray spectrum. The QSOIIs from \citet{toz06} are restricted
to those AGN with spectroscopic redshifts and no broad lines from
\citet{szo04}, leaving only 13 sources. Due to the low number of
detected photons, $< 100$ total counts, some of the X-ray spectral
fits have too much scatter for inclusion in table 1. In
particular, the modest absorption ($N_{H} \approx 2\times
10^{23}\mathrm{cm}^{-2}$) source, CDFS-112 at $z=2.9$ and the
highly absorbed sources CDFS -531, CDFS - 263, CDFS-268 have X-ray
data that are not adequately fit by the models of \citet{toz06}.
\begin{figure}
\plotone{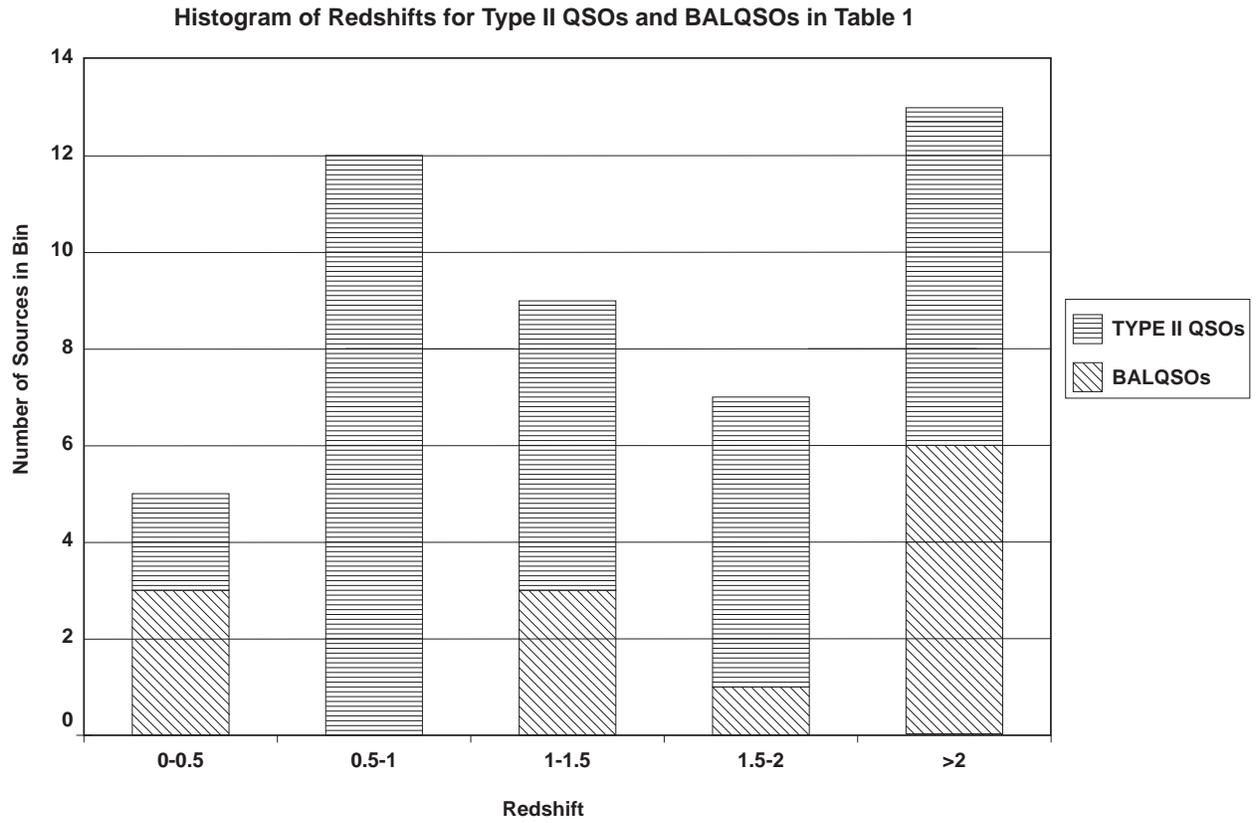}
    \caption{A histogram of the distributions of redshifts of
    BALQSOs and QSOIIs in the sample in table 1}
\end{figure}
\subsection{Selection Effects in BALQSO Sample}The BALQSO X-ray observations typically
select the brightest X-ray objects to insure the best chance of
successful detection of sufficient counts to determine a
reasonable spectral fit. The ASCA deep survey of BALQSOs in
\citet{gal99} used $>20$ ks exposures to place lower limits on
$N_{H}$, indicating that 5 BALQSOs sources are likely candidates
to be Compton thick with deep Chandra or XMM observations, IRAS
0759+6508, LBQS 2111-4335, PG 1700+518, PG 0043+034 and 0043+008.
There are no followup observations of these sources by any of
these authors with Chandra or XMM in the literature. Yet some
members of this research team chose to observe PG 2112+059 which
was known to be X-ray bright from previous ROSAT and ASCA
observations \citep{gal01} and SBS 1547+517 which was known to be
X-ray bright from Beppo SAX observations in \citet{bra99}.
Similarly \citet{gre01} performed a snapshot survey of BALQSOs
with Chandra and some of the research team chose to study the two
brightest sources UM 425, SBS 1542+541 and the brightest ROSAT
BALQSO from \citet{gre96}, 1246-0542 with deeper observations.
There are are no published Chandra or XMM data on the
preponderance of weaker sources in the snapshot survey of
\citet{gre01}. The point of this is merely to illustrate a strong
selection effect produced five of the six BALQSO values of
$N_{H}\leq 10^{23}\mathrm{cm}^{-2}$ in table 1.
\par Another selection effect can be seen from the redshift
distribution in the histogram of figure 2. The BALQSOs are
selected by optical considerations which usually is the shifting
of the CIV absorption trough into the visible band. Thus, these
are predominantly sources with $z>1.5$. The QSOIIs are mostly
X-ray selected so there are more $z<1$ sources as a result of the
flux limitations of the deep surveys. It is quite possible that
selecting BALQSOs from flux limited X-ray samples would alter the
results of this paper. However, the preponderance of high redshift
BALQSOs in table 1 does not explain the anomalously large column
densities of BALQSOs relative to QSOIIs because most of largest
BALQSO column densities in table 1 are found at low redshift.
\subsection{Can Column Densities be Compared?}
The QSOII sample is not intended to be representative of the
distribution of $N_{H}$ for the entire hidden quasar population.
However, the QSOII sources found by the methods above and
tabulated in table 1 are sufficient for the purposes of comparison
with BALQSOs within the context of the BAL equatorial wind
extension of the standard model. In particular, this (or any)
sample of QSOIIs, by definition, not only has an obscured blue
side of the UV BEL region (the near side of the equatorial wind
model), as for BAL QSOS, but the red side (far side) of the UV BEL
region as well as the central optical/UV continuum are also
obscured. Thus, the ubiquity of BAL winds in all radio quiet QSOs
within this extension of the standard model, implies that any
subsample of QSOIIs should be more obscured than BALQSOs on
average. In particular, a BALQSO has two dominant components to
the absorption column density, $\Sigma_{\mathrm{shielding}}$,
which is a filter that absorbs X-rays from the central continuum
that allows the ionization state to be low enough so that
Lithium-like atoms can form in the BAL wind zone farther out,
$\sim 10^{16}\mathrm{cm}- 10^{17}\mathrm{cm}$ away \citep{mur95}.
Secondly there is the absorption in the wind itself,
$\Sigma_{\mathrm{wind}} \ll \Sigma_{\mathrm{shielding}}$ which is
virtually negligible, $\Sigma_{BALQSO}=\Sigma_{\mathrm{wind}}+
\Sigma_{\mathrm{shielding}}\approx \Sigma_{\mathrm{shielding}} $
\citep{mur95}. The shielding gas might actually just be the dense
base of the BAL wind \citep{pro00,pro04}.
\par The QSOIIs have an
additional absorbtion component from the intervening dusty torus
$\Sigma_{\mathrm{torus}}\gg \Sigma_{\mathrm{wind}}$. A line of
sight through the torus must also pass through the shielding gas
on the way to the central X-ray source. This is a very important
constraint on the model. The shielding gas has its highest density
just above the accretion disk in both the theoretical treatment of
\citet{mur95} and in the numerical results of \citet{pro04}. This
is simply a consequence of the mass conservation law as the
shielding gas has its source in a low velocity outflow emerging
from the disk atmosphere or corona. To be precise, the type II
lines on sight should encounter a larger
$\Sigma_{\mathrm{shielding}}$ than BALQSO lines of sight in the
equatorial wind model. Thus, the typical QSOII absorbing column
should exceed the typical BALQSO absorbing column,
$\Sigma_{\mathrm{type
II}}=\Sigma_{\mathrm{shielding}}+\Sigma_{\mathrm{wind}}+
\Sigma_{\mathrm{torus}}>\Sigma_{\mathrm{BALQSO}}=\Sigma_{\mathrm{shielding}}+\Sigma_{\mathrm{wind}}$,
by an amount roughly equal to the typical absorption column
through the dusty torus.
\par It is not trivial to argue that the type II lines of sight
might miss the shielding gas, i.e. the shielding gas is suspended
at high latitude above the disk. This is a serious modification to
the model for two reasons. First, it changes the source of the
shielding gas to an unknown entity which is now adhoc. Secondly,
coronal X-rays could now pass underneath the shielding gas and
over-ionize the base of the wind zone. This would prevent resonant
absorption from transferring sufficient moment to launch the BAL
wind vertically away from the disk, its mass source \citep{mur95}.
Such a modification is a significant departure from the equatorial
wind model that is considered credible largely because of its physical
explanation of the X-ray shielding mechanism.
\begin{figure}
\plotone{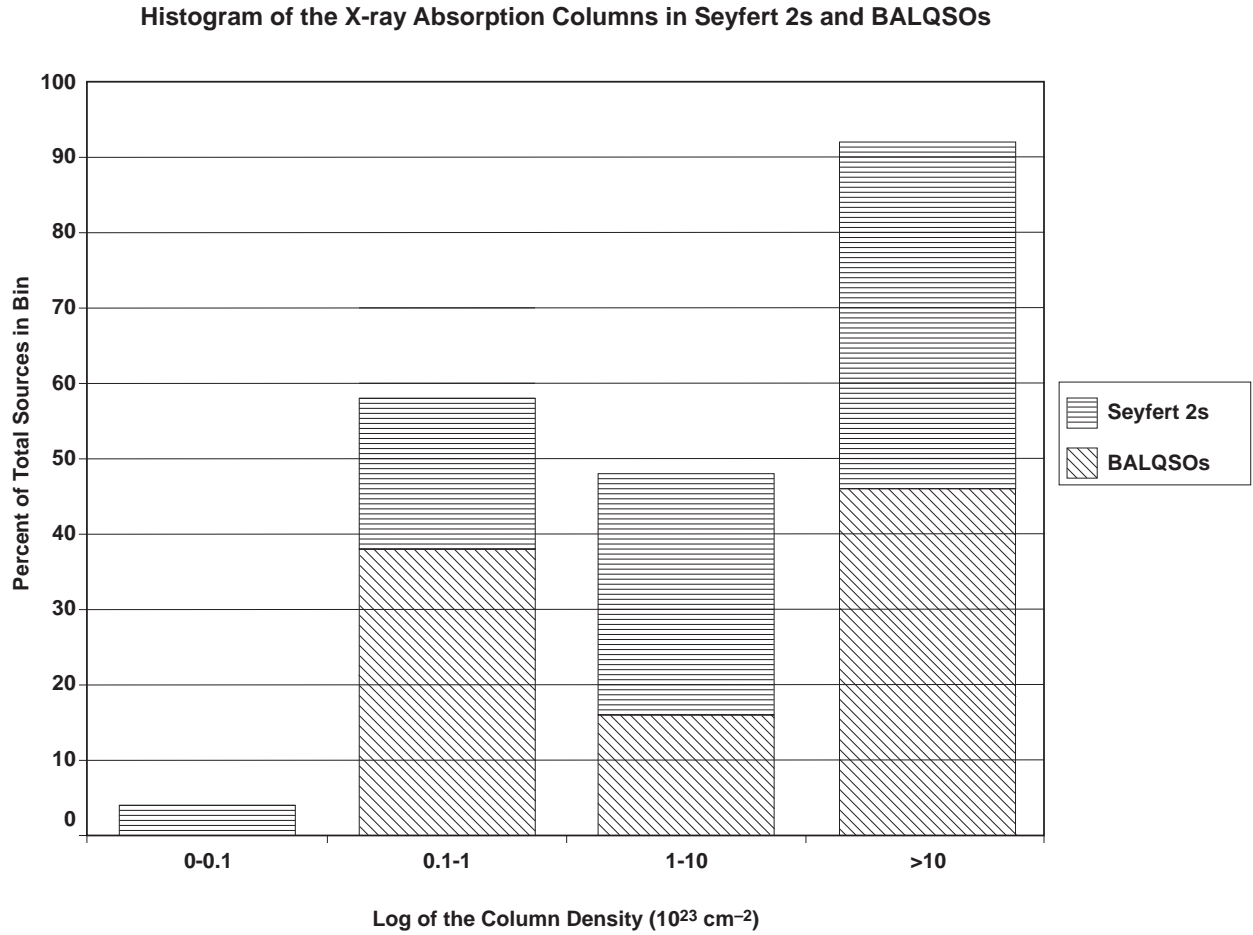}
    \caption{A histogram of the distributions of the hydrogen column densities of BALQSOs with deep X-ray observations
    compared to those for Seyfert 2 galaxies
    from \citet{wil02}.}
\end{figure}
\section{Interpretation of the Results} The data in table 1 is summarized
in the histogram in figure 1. The data does not support the
hypothesis that QSOIIs have larger X-ray attenuating columns than
BALQSOs (as expected from the equatorial wind model). In fact a
Mann-Whitney test shows that the BALQSOs have larger values of
$N_{H}$ than those of the QSOIIs in table 1 at the $99.8\%$
significance level. The result is also true above the $97.8\%$
significance level in Kolmogorov-Smirnov two sample test. One
would have to add 11 Compton thick QSOII sources to the sample in
order for the significance level to drop below $95\%$ in the
Mann-Whitney test. The discussion in the previous section shows
that there are some strong selection effects as well as a small
sample size that could make a general interpretation of the
results in table 1 misleading. However, there has certainly been
no effort to select for the most attenuated BALQSOs, but actually
some effort to the contrary. For example, 5 of the 13 of the
BALQSOs were selected for deep observation because they were
particularly X-ray bright for BALQSOs (see section 3.3). Removing
the 5 BALQSOs chosen for deep observation by X-ray brightness
increases the significance to $> 99.99\%$ by a Mann-Whitney test.
\par The primary question is how many Compton thick QSOIIs were
missed by selection effects. It seems that as more deep X-ray
spectra of QSOIIs become available that the statistical
significance of the  result might well diminish or disappear due
the the inclusion of many weaker Compton thick sources. As noted
above, only 11 more Compton thick QSOIIs would remove the
statistical significance of the result. In order to investigate
this possibility, one can compare the absorption columns in nearby
Seyfert 2 galaxies to see if it reasonable to expect a large
fraction of Compton thick QSOIIs. This exercise is pragmatically
motivated since the local Seyfert 2 population is easier to study
in flux limited samples which can still detect nearby Compton
thick sources. However, this expedience goes directly against the
admonishment of section 2 that Seyfert galaxies do not have a
direct relationship to BALQSOs, but it can yield some "ballpark"
type of information. Figure 3 is a histogram of the distribution
of Seyfert 2 galaxy hydrogen absorption columns given in
\citet{wil02} compared to the BALQSO sample presented here. The
data is normalized so each bin represents the percent of that
sample that is in the corresponding X-ray absorption column range.
The distribution for Seyfert 2s and BALQSOs are incredibly
similar. We do not compare them statistically because there is no
physical basis for Seyferts to be the parent population of BALQSOs
as discussed in section 2. However, the Seyfert 2 distribution
indicates that $\approx45\%$ of QSOIIs might be Compton thick
which seems consistent with the results of \citet{toz06}. Thus, a
deep survey should produce enough Compton thick sources to remove
the statistical significance of the result that BALQSOs have larger
values of $N_{H}$ than those of the QSOIIs in table 1.
\par However, it is not the claim of this article that BALQSOs have larger values of
$N_{H}$ than those of the QSOIIs. The claim of this article is
that the observations do not support the notion that $N_{H}$ is
larger in QSOIIs than BALQSOs. In particular, the Seyfert 2
results in figure 3 do not suggest that deeper X-ray observations
of QSOIIs will reverse the trend in figure 1 (especially if more
X-ray weak BALQSOs are observed), This is primarily a consequence
of the significant number of QSOIIs with $N_{H}\leq
10^{23}\mathrm{cm}^{-2}$ seen in table 1 and in the sample of
\citet{toz06}, compared to the majority of BALQSOs with $N_{H}\geq
10^{23}\mathrm{cm}^{-2}$. Quantitatively speaking, one would have
to add 145 more QSOIIs to the sample in table 1 that are Compton
thick in order to reverse the trend and attain a result that the
QSOII values of $N_{H}$ are larger than the BALQSO values of
$N_{H}$ at a marginal level of statistical significance. This
would correspond to $83\%$ of the QSOIIs being Compton thick which
is not supported by the Seyfert 2 data in figure 3. As discussed in section 3.3, if the two
populations are statistically indistinguishable then it is still a
conundrum for the equatorial wind model, the torus must be a more
significant absorber than the BAL wind forming its upper skin.
\par This statistical study indicates that some or all of the
BALQSOs contain an additional, large, absorbing region that is not
encountered by lines of sight (near the equatorial plane) typical
of QSOIIs. The extra absorber would have an average column density
$\sim \Sigma_{\mathrm{torus}}-\Sigma_{\mathrm{wind}}\sim
\Sigma_{\mathrm{torus}}$. Two possible explanations of the extra
absorber can be found in the empirical hollow cone outflow model
of QSOs in \citet{elv00} and the analytically calculated polar
wind models of \citet{pun99,pun00}. In fact, it is claimed that a
polar BAL wind has been directly observed in MRK 231
\citet{pun05}. Another 6-8 polar BALQSOs have been found in
\citet{zho06}. If the X-ray absorber and the BAL wind are along
the polar axis in many BALQSOs (not necessarily all), then type II
lines of sight would never encounter these large absorption
columns.
\par It is interesting that 4 of the 5 highest column densities for
BALQSOs show low ionization UV absorption troughs, MRK 231,
PHL5200, IRAS 07598+6508 and 1413+117. If these sources are
removed from the BALQSO sample then the BALQSOs have absorption
columns that are larger than QSOIIs at only a marginal level of
significance by a Mann-Whitney test, $95.4\%$. This might suggest
that the extra absorber is more prominent in BALQSOs with low
ionization absorption. However, one can not justify removing these
sources from the sample since according to the equatorial wind
model, the low ionization sources are merely those BAL winds
viewed even closer to the torus (the lowest BAL latitudes)
\citep{mur95}. So technically, a sample of pure low ionization
BALQSOs should still show less absorption than a QSOII sample.
\section{Discussion} This article is the first attempt to compare
the absorption column densities of QSOIIs (completely enshrouded
QSOs) and BALQSOs (partially shrouded QSOs). It is claimed that
this comparison is an important clue for understanding the
fundamental underlying physical configuration that constitutes a
QSO. The preliminary results presented here seem to suggest that
the subpopulation of QSOIIs are not more absorbed than BALQSOs in
general. There is a strong indication that a significant fraction
of the BALQSO population is not described by an equatorial BAL
wind, the X-ray absorbing gas is not concentrated at low
latitudes. Hopefully, this discussion will motivate future X-ray
observations of large samples of objects with minimal selection
effects. One way to choose the objects might be with a MID IR flux
limited sample created from deep Spitzer observations.


\begin{thebibliography}{}
\bibitem[Aldcroft and Green(2003)]{ald03} Aldcroft, T.L., Green, P.J., 2003 ApJ \textbf{592} 710
\bibitem[Alexander et al (2005)]{ale05}Alexander, D.M. et al 2005, ApJ \textbf{632}
736
\bibitem[Antonucci(1993)]{ant93} Antonucci, R.J. 1993,
  Annu. Rev. Astron. Astrophys. \textbf{31} 473
\bibitem[Belsole et al(2006)]{bel06} E. Belsole , D. M. Worrall , M.J. Hardcastle astro-ph/0511606 to appear in MNRAS
\bibitem[Braito et al(2004)]{bra04}Braito, V. et al A \& A 2004 \textbf{420}
79
\bibitem[Brandt et al (1999)]{bra99}Brandt, W.N., et al 1999, ApJL \textbf{525}
69
\bibitem[Brandt and Schulz(2001)]{bra01}Brandt, W.N., Schulz, N. 2001, ApJ \textbf{544}
123
\bibitem[Brinkmann et al(2002)]{bri02}Brinkmann, W., Ferrero, E., Gliozzi, M. 2002, A \& A
\textbf{385} 31
\bibitem[Cacciangia et al(2004)]{cac04} Cacciangia, A. et al 2004, A \& A
\textbf{416} 901
\bibitem[Clavel et al(2006)]{cla06}Clavel, J. Schartel, N., Tomas, L.  2006 astro-ph/0509432 to appear in A \& A
\bibitem[Chapman et al (2005)]{cha05}Chapman, S., et al 2005, ApJ \textbf{622}
772
\bibitem[Chartas et al(2002)]{cha02}Chartas, G., Brandt, W. N., Gallagher, S. C. and Garmire, G. P. 2002, ApJ \textbf{579} 169
\bibitem[Chartas et al(2003)]{cha03}Chartas, G., Brandt, W. N., Gallagher, S. C. 2003, ApJ \textbf{595}
 85
\bibitem[Chartas et al(2004)]{cha04}Chartas, G. Eracleous, M. Agol, E. and Gallagher, S. 2004 ApJ \textbf{606} 78
\bibitem[Elvis(2000)]{elv00} Elvis, M. 2000 ApJ \textbf{545} 63
\bibitem[Evans et al(2006)]{eva06}Evans, D., Worrall,D., Hardcastle,M., Kraft,R.,Birkinshaw,
M. 2006 astro-ph/0512600 to appear in ApJ
\bibitem[Gallagher et al(1999)]{gal99}Gallagher, S et al
ApJ 1999 \textbf{519} 549
\bibitem[Gallagher et al(2001)]{gal01}Gallagher, S et al
ApJ 2001 \textbf{546} 745
\bibitem[Gallagher et al(2002)]{gal02}Gallagher, S, Brandt, W.N., Chartas, G., Gamire, G.P.
ApJ 2002 \textbf{567} 37
\bibitem[Gallagher et al(2004)]{gal04}Gallagher, S. et al 2004 ApJ \textbf{603} 425
\bibitem[Gallagher et al(2006)]{gal06}Gallagher, S. et al 2006 to
appear in ApJ astro-ph/0602550
\bibitem[Georgantopoulos et al(2006)]{geo06}Georgantopoulos, I.  et al
2006, MNRAS in press astro-ph/0601245
\bibitem[Green and Mathur(1996)]{gre96}Green, P. and Mathur, S. 1996 ApJ \textbf{462} 637
\bibitem[Green et al(2001)]{gre01}Green, P. et al 2001 ApJ \textbf{558}
109
\bibitem[Grupe et al(2003)]{gru03}Grupe, D., Mathur, S., Elvis, M., 2003 AJ \textbf{126} 1159
\bibitem[Hasinger et al (2002)]{has02}Hasinger, G., Schartel., N. Komossa, S., 2002 ApJ
\textbf{573} L77
\bibitem[Hewett and Foltz (2003)]{hew03}Hewett, P. and Foltz, C., 2003 AJ
\textbf{125} 1784
\bibitem[Imanishi and Terashima (2004)]{ima04}Imanishi, M. and Terashima, Y., 2004 AJ
\textbf{127} 758
\bibitem[Iwasawa et al (2001)]{iwa01}Iwasawa, K, Fabian, C., Ettori, S. 2001
MNRAS \textbf{321} L15
\bibitem[Mathur et al(2000)]{mat00}Mathur, M. et al ApJ 2000 \textbf{533}
79
\bibitem[Mathur et al(2001)]{mat01}Mathur, S. et al ApJ 2001
\textbf{551}13
\bibitem[Morgan et al(2001)]{mor01}Morgan, N. et al  ApJ 2001 \textbf{555} 1
\bibitem[Murray et al(1995)]{mur95} Murray, N. et al 1995, ApJ \textbf{451}
498
\bibitem[Norman et al(2002)]{nor02} Norman, C. et al 2002 ApJ \textbf{571} 218
498
\bibitem[Oshima et al(2001)]{osh01} Oshima, T. et al 2001 ApJ \textbf{563}
103
\bibitem[Perola et al(2004)]{per04} Perola, G, et al 2004, A \& A
\textbf{421} 491
\bibitem[Polletta et al(2006)]{pol06}Poletta et al 2006
astro-ph/0602228 to appear in ApJ
\bibitem[Proga et al(2000)]{pro00}Proga, D., Stone, J., Kallman, T. ApJ 2000 \textbf{543}
686
\bibitem[Proga and Kallman(2004)]{pro04}Proga, Kallman, T. ApJ 2004 \textbf{616}
688
\bibitem[Ptak et al(2006)]{pta06}Ptak, A. et al 2006 ApJ \textbf{637}
137
\bibitem[Punsly(1999a)]{pun99}Punsly, B. 1999, ApJ \textbf{527} 609
\bibitem[Punsly(1999b)]{pun00}Punsly, B. 1999, ApJ \textbf{527} 624
\bibitem[Punsly and Lipari (2005)]{pun05}Punsly, B. 2005, ApJL \textbf{623}
101
\bibitem[Sabra et al(2001)]{sab01}Sabra, B. and Hamman, M. 2001 ApJ \textbf{563} 555
\bibitem[Severgnini et al(2006)]{sev06}Severgnini, P. et al 2006
astro-ph/0602486 to appear in A \& A
\bibitem[Shemmer et al(2006)]{she06}Shemmer, O. astro-ph/0509146 2006 ApJ in press
\bibitem[Sturm et al(2006)]{stu06}Sturm, E. et al 2006 astro-ph/0601204 to appear in ApJ
\bibitem[Szokoly et al(2004)]{szo04}Szokoly, G. et al 2004 ApJS \textbf{155}
271
\bibitem[Tozzi et al(2006)]{toz06}Tozzi, P. et al 2006 astro-ph/0602127
to appear in A \& A
\bibitem[Weymann (1997)]{wey97}Weymann, R. 1997 in ASP Conf. Ser.
128, \textbf{Mass Ejection from Active Nuclei} ed, N.Arav, I.
Shlosman and R.J. Weymann (San Francisco: ASP) 3
\bibitem[Wilkes et al(2002)]{wil02}Wilkes, B.. et al 2002 ApJL \textbf{564}
L65
\bibitem[Zhou et al (2006)]{zho06}Zhou, H. et al 2006, ApJ \textbf{639} 716
\end{thebibliography}
\end{document}